# Molecular Photodissociation in the Neutral Envelopes of Planetary Nebulae

G. T. Gussie

Department of Physics, University of Tasmania

GPO Box 252C

Hobart, Tasmania, Australia 7001

internet: Grant.Gussie@phys.utas.edu.au

A. R. Taylor

Department of Physics and Astronomy, The University of Calgary

2500 University Drive N.W.

Calgary, Alberta, Canada T2N 1N4

internet: russ@ras.ucalgary.ca

P. E. Dewdney and R. S. Roger

Dominion Radio Astrophysical Observatory

Herzberg Institute of Astrophysics, National Research Council

PO Box 248 Penticton, BC, Canada V2A 6K3

internet: ped@drao.nrc.ca

Abstract

The dissociation and subsequent ionisation of molecular material in a detached, spherically symmetric stellar wind is modelled as a star evolves from the post-asymptotic giant branch phase to a planetary nebula (*i.e.* from a stellar effective temperature of $T_{eff}$ = 6000 K to $T_{eff}$ = 33000 K). The model is compared to the observed mass and distribution of ionised and atomic hydrogen in the planetary nebula IC 418. It is found that the mass of circumnebular atomic hydrogen observed in IC 418. can easily be produced by the dissociation of originally molecular material by the ultraviolet radiation of the central star. However, the observed spatial distribution of the gas is not accurately recreated by the current model.

I. INTRODUCTION

It is now well established that planetary nebulae are the ionised remnants of originally neutral material that was lost as a stellar wind during a star's asymptotic-giant-branch (AGB) or post-AGB evolution. It is also well established that the ionisation of the stellar wind is incomplete in many planetary nebulae, as is shown by the discovery of molecular and atomic envelopes surrounding a significant number of these objects (e.g. Zuckerman & Gatley 1988; Huggins & Healy 1989; Taylor, Gussie, & Pottasch 1990; hereafter TGP).

However, the chemical and physical evolution of the neutral envelopes of planetary nebulae are still poorly understood. For example, it is as yet not clear whether the observed circumnebular atomic hydrogen (HI) results from an originally atomic stellar wind or was an originally molecular wind that has since been dissociated. If one assumed that the atomic gas was once molecular, it would then be unclear whether the atoms were formed from molecular dissociation caused by collisions (for example by the passage of a shock front) or whether molecular photodissociation predominated. The relative importance of the radiation of the central star and interstellar radiation in the process of molecular photodissociation is also unclear at the present time.

An understanding of the formation and evolution of the HI envelopes of planetary nebulae can only come after thorough observational study. A relatively well-observed planetary nebula that is currently known to possess HI is IC 418 (Taylor & Pottasch 1987). This nebula has been previously mapped in $\lambda = 21$ cm HI emission by Taylor, Gussie, & Goss (1989; hereafter TGG).

It is believed to be unlikely that the HI of IC 418 was once molecular material that has been dissociated by the passage of a strong shock. A shock with a velocity of $\geq 24$ km s$^{-1}$ could have dissociated the original $H_2$ molecules (Kwan 1977), but it is not easily seen how such a shock could be generated within the neutral envelope. The expansion velocity of the IC 418's ionised gas is very similar to that of the surrounding envelope (TGP), indicating that the nebula's ionised and neutral components are not undergoing violent collision. Shock excitation of $H_2$ would also be expected to produce observable vibrational-rotational transitions of $H_2$, which are not seen in IC 418 (Webster et al. 1988). The atomic shell is also very much larger than the ionised nebula; extending to a radius of $\theta_{HI} \geq 90"$ (TGG) as compared to the ionised radius of $\theta_{ion} \approx 7.5"$ (Balick 1987). The atomic shell also has an apparently smooth, inverse-square-law density gradient such as would be expected for an un-shocked stellar wind (TGG). It therefore appears unlikely that the atomic envelope of IC 418 represents post-shock gas.

It also appears unlikely that the circumnebular HI of IC 418 is the remnant of a period of atomic mass loss during the precursor star's AGB evolution. There is no observational evidence of significant atomic mass loss occurring in those stars presently on the AGB (e.g. Knapp & Bowers 1983). Therefore, unless the AGB precursor of IC 418 was an unusual star, the hypothetical period of atomic mass loss would likely have been very brief and must have occurred some time during the post-AGB evolution to the planetary nebula phase. The dynamic age of the atomic material would then be small compared to AGB evolutionary time scales. However, the dynamical age of the atomic gas cloud is $\geq 37000$ yr (TGG), which is on the same order as the time scale of evolution from the AGB to planetary nebulae (Schönberner 1983). The stellar wind that created the atomic material would therefore likely have left the star while it was still on the AGB, and would have necessarily continued throughout most of the star's subsequent evolution to become a planetary nebula. It is therefore probable that the atomic wind material seen around IC 418 was originally molecular, as are the stellar winds of observed in those stars currently in the AGB or proto-planetary nebulae phases.

It has been shown (TGG) that the $\geq 37000$ yr age of the circumnebular envelope of IC 418 is sufficient for the observed mass of HI to have been formed from an originally molecular stellar wind through photodissociation by the interstellar radiation field. However, subsequent observations of other young planetary nebulae by TGP have shown that interstellar dissociation could have played only a very minor role in the formation of HI around these objects. Furthermore, the HI emission of the planetary nebula BD +30° 3639 has been shown to be confined to within an angular radius of $\theta_{HI} \leq 24"$ (Likkel et al. 1992), which is inconsistent with the extended HI distribution that would be expected if its HI were formed by interstellar photodissociation. It is however probable that IC 418's envelope is quite different than that of BD+30° 3639, which is known to have a significant molecular component (Bachiller et al. 1991, Graham et al. 1993). In contrast, IC 418 is not observed to possess either a CO envelope (Huggins & Healy 1989) or shock-excited $H_2$ (Webster et al. 1988). We therefore suspect that IC 418 is significantly less massive and somewhat more evolved than BD+30° 3639 (Likkel et al 1992). Therefore, if the HI observed in IC 418 was formed primarily by interstellar photodissociation, this nebula's atomic envelope would seem to be unique among those HI envelopes currently known.

The purpose of this paper is to investigate the alternative hypothesis that the observed circumnebular atomic hydrogen in IC 418 had been formed primarily by the photodissociation of an originally molecular stellar wind by the ultraviolet radiation from the central star.

II. OBSERVATIONS

$\lambda$ = 21 cm observations were made over 8 hours with the Very Large Array (VLA) in D configuration. Twelve hours of supplementary observations were also made with the 1.5B configuration of the Australia Telescope National Facility Compact Array (ATCA). The ATCA's 1.5B configuration is a relatively compact configuration, with baseline lengths ranging from 31 m to 4301 m. Twenty-six VLA antennae were used, as were all 6 ATCA antennae. The observing dates were 12 January 1990 for the VLA observations and 22 March 1993 for the ATCA observations.

The $uv$ visibility data from the two telescopes were separately edited, calibrated, and self-calibrated within the Astronomical Image Processing System (AIPS). Spectral line maps were then made from the VLA visibilities alone. Complementary maps were also made with the VLA visibilities augmented by data from the ATCA observations. In order to add the ATCA data to the VLA data, the ATCA data were first shifted to the same Döppler-corrected velocity scale as the VLA data and then averaged to the VLA's lower velocity resolution. Velocity averaging and combination of the $uv$ data sets was done within the data reduction program MIRIAD. Both the VLA data and the ATCA-augmented data have 127 spectral line channels, with 2.56 km s$^{-1}$ velocity resolution, and with the central channel at the local standard of rest velocity of the nebula; $v_{LSR}$ = 43.4 km s$^{-1}$ (Schneider et al. 1983).

Augmenting the VLA's $uv$ data with the ATCA visibilities increased the number of data points by ≈ 20%. The nebula IC 418 is a low declination object ($\delta$ = -12°), making the projected baselines on the east/west linear array of the ATCA very short. The ATCA visibilities are consequently concentrated toward the center of the $uv$ plane (< 100 m), and therefore serve to degrade the maps' spatial resolution (especially in declination). By doing so, it was hoped that maximum sensitivity to extended emission can be obtained. Such sensitivity is desired because observations by TGG have shown that the HI envelope of IC 418 is quite large and likely possesses an inverse-square-law density gradient. TGG could therefore obtain only a lower limit to the extent of HI emission. The synthesised beam size of the VLA-only maps is 50" × 55", while the addition of the ATCA visibilities increase the beam size to 55" × 77".

Consequently the new observations have, in comparison to the original mapping experiment of TGG, lower spatial resolution; but have improved sensitivity to extended emission, improved $uv$ coverage, and also have twice the velocity resolution.

Maps of the HI envelope were made by subtraction of the ionised gas' continuum radiation within the $uv$ (Fourier transform) domain with the AIPS task UVLSF. Mapping and cleaning of the continuum-subtracted visibilities was performed by the task MX. Flux in the continuum-subtracted maps was measured with the task IRING, which measures flux in concentric annuli. By centering these annuli on the position of the continuum source the cumulative flux contained within a beam-sized annulus and its interior was determined. The source flux is defined to be the cumulative flux within the beam-sized annulus, minus an average background flux. The background flux was determined from the azimuthally averaged flux of a single annulus exterior to the beam-sized annulus. The exterior annulus has a sufficient width ($dr$) to give it the same surface area as the beam, allowing the background flux to be removed from the source by simple subtraction.

The five spectral channels displaying net HI emission have been averaged to determine the spatial distribution of circumnebular HI in IC 418. The resulting map is shown in figure 1. An integrated HI line flux of $S_{HI}$ = 42±5 mJy is observed. Maps made with the VLA data augmented with ATCA data shown no evidence of additional extended emission over the VLA-only map shown in figure 2. The ATCA-augmented maps are also (unfortunately) barely resolved in declination and so have not been used for the subsequent analysis despite their improved signal-to-noise ratio over the VLA-only maps.

Despite the successful use of self-calibration and the excellent $uv$ coverage provided by the two telescopes, there remains some negative troughs in the final maps at the 0.2 mJy level. The shape of the outer contours of the HI map is affected by these negative bowls. However, the dependability of the analysis of the image of circumnebular HI in IC 418 is greatly increased if the total flux is considered, rather than the HI emission's surface brightness. To determine total flux as a function of angular radius, the AIPS task IRING was used to first determine the azimuthally-averaged surface brightness (measured in mJy beam$^{-1}$) within concentric annuli centered on the position of the ionised nebula. The total flux contained in each annulus was then determined by multiplying the measured surface brightness by the area of the annulus. This results in a radial flux profile, shown in figure 3.

Error bars for the radial flux profile are difficult to estimate because of the aforementioned negative troughs in the maps. The base level of the maps is quite smooth over the diameter of the small annuli, but is variable over the diameter of the large annuli. The error bars were determined by measuring the RMS noise level of a large patch of the CLEANed map that lacked HI emission or a continuum source. This patch was made large enough to include two negative troughs, which increased the measured variation in the pixel values greatly. The "noise" level that was measured is therefore a combination of instrumental effects (troughs) and thermal noise. The error bars are calculated from the measured RMS noise by multiplying by the area of each annulus, resulting in error bars that increase in size with radius.

III. EMPIRICAL MODEL OF THE CIRCUMNEBULAR ENVELOPE OF IC 418

To understand the morphology and evolution of the nebula, the circumnebular atomic hydrogen is modelled empirically by simultaneously fitting the $\lambda = 21\,\text{cm}$ spectrum (figures 1 & 5a) and the radial flux profile (figures 3 & 5b) with four proposed spherically-symmetric morphologies.

The empirical determination of the HI mass, density power law, temperature, and expansion velocity from the simultaneous fitting of the radial flux profile and the $\lambda = 21\,\text{cm}$ spectrum is described in TGG. Four model morphologies are considered in the present modelling, as were also considered by TGG. The model morphologies are: a constant expansion velocity, constant density ($n \propto r^0$) model ; a constant expansion velocity, constant mass-loss rate model ($n \propto r^{-2}$); a constant expansion velocity, increasing mass-loss rate model ($n \propto r^{-3}$); and a radially increasing expansion velocity, constant mass-loss rate model ($n \propto r^{-3}$). The parameters of all four models are varied until the best-fit parameters for each are found. However, the empirical model used (shown in figure 4) is somewhat different than the original TGG model in that it allows the option of *two* HI shells, rather than the single HI shell model of TGG. It is thought that an inner HI shell would result from photodissociation by ultraviolet radiation from the central star while a hypothetical outer HI shell would form from interstellar ultraviolet radiation. The presence of an undissociated molecular region between two atomic shells is therefore permitted (but not required). Another change made to the original TGG model is that TGG allowed the inner edge of the HI shell to be a free parameter. The justification for this was that an observation reported by Reay & Worswick (1979) suggested that IC 418 possesses an extended ionised halo. If this were true, the inner edge of the HI shell could lie well outside the bright ionised shell that is observable as a radio continuum source. However, subsequent observations by Phillips, Riera, & Mampaso (1990) have shown that the extended H$\alpha$ halo is most likely scattered radiation from circumnebular dust (which was also observed directly as infrared emission). There is consequently no strong evidence of extended ionised gas around IC 418 at the present time. It is therefore assumed in the current model that the inner edge of the inner HI shell is coincident with the outer edge of the radio-bright ionised shell.

The homogeneous ($n \propto r^0$) shell produced by the constant expansion velocity, constant gas density model morphology is the least satisfactory of the four models (figure 5). The observed radial flux profile can not be fit to within the error bars with any set of parameters. Any model that correctly reproduces the angular offset of the peak flux ($\Delta\theta$ = 30" to $\Delta\theta$ = 40") results in a severely underestimated flux at greater radii. The flux at smaller angles, notably $\Delta\theta$ = 10" to $\Delta\theta$ = 20", is also underestimated. This model also fails to adequately reproduce the shape of the $\lambda$ = 21 cm spectrum.

It is also unclear how easily a constant expansion-velocity, constant gas density morphology can be physically realized. A *thin* homogeneous shell could be created by a shock propagating through a larger, low-density HI complex. However, as stated, the large size of the HI cloud of IC 418 makes it unlikely that it is entirely composed of post-shock gas. One must therefore postulate an ad hoc, time-variable mass-loss rate to create a homogeneous shell. The constant expansion velocity, constant gas density model therefore does not seem to be a reasonable model of the circumnebular HI of IC 418.

The two $n \propto r^{-3}$ density power law models produce acceptable fits to the spectrum, but may be too "peaky" to fit the radial flux profile as satisfactorily as the $n \propto r^{-2}$ model. The $n \propto r^{-2}$ density power law model produces the best simultaneous fit to the radial flux profile and the spectrum. It is therefore accepted that the simple $n \propto r^{-2}$ density power law — created by a constant expansion velocity, constant mass-loss rate stellar wind — is an acceptable model of the HI envelope of IC 418.

The $n \propto r^{-2}$ model morphology and the two $n \propto r^{-3}$ model morphologies all have two separate HI shells in their "best-fit" parameter sets. The best-fit models of these morphologies can therefore be described as a massive outer shell at an angular radius of $\geq$ 25" and a very thin, inner shell near 7.5". The inner shell is required to increase the models' absorption line depth relative to the height of the model emission line, and also to increase the low-radius flux of the radial flux profile. However, the empirical models with only one HI shell which have an inner edge that is allowed to vary (as were the models of TGG) also produce acceptable fits to the data. The data therefore do not show conclusively that a double HI shell model of IC 418's envelope is required. However all single-shell model fits require that the inner edge of the HI shell be at $\approx 25"$, which is well outside the observed extent of the ionised gas and consequently inconsistent with the proposed hypothesis of photodissociation by the central star. Therefore there does not seem to be any conclusive evidence of a stellar photodissociation region in IC 418, and if one does exist, it would appear to very thin.

It is now asked whether the thin stellar photodissociation region of the "best-fit" empirical model is what would be expected from the photodissociation of a molecular wind by the central star. The growth of both the photoionised region and the photodissociated region are modelled to determine the predicted size and mass of the HI envelope as a function of time under the hypothesis of HI formation by photodissociation of molecular gas by ultraviolet radiation from the central star.

IV. EVOLUTIONARY MODEL OF THE IONISED REGION

The formation and evolution of the ionised regions of planetary nebulae is currently a topic of intense interest. Within the last decade, the interacting stellar winds model of Kwok, Purton, & Fitzgerald (1978) has gained very wide acceptance. In this model, the slow stellar wind of the nebula's AGB precursor star is overtaken by a fast wind from the nebula's central stellar remnant. The shocked interaction between the two winds results in the dense ionised gas shell seen in planetary nebulae. However, there is some question as to whether the interaction of stellar winds is important to the kinematics of the very youngest planetary nebulae. Hydrodynamical calculations of planetary nebula evolution based on the interacting stellar winds model (e.g. Mellema, Eulderink, & Icke 1991) predict that the ionised gas should undergo a significant acceleration to velocities $\gtrsim 10$ km s$^{-1}$ higher than the expansion velocities seen in the circumstellar envelopes of AGB stars. It has also been found that such acceleration should occur quite early in the ionised nebula's evolution, while the radius of the ionised nebula is only $R_{ion} \leq 0.05\,\mathrm{pc}$ (Kahn & Breitschwerdt 1990). However, it is found that the smallest planetary nebulae (with ionised linear radii of $R_{ion} \leq 0.05\,\mathrm{pc}$) show no evidence of such ionised gas acceleration, although it is often quite evident in larger nebulae (Gussie & Taylor 1994; hereafter GT).

An alternative to the interacting stellar winds model that may be appropriate (or at least adequate) for very small planetary nebulae is the photoionised wind model (Taylor, Pottasch, & Zhang 1987; hereafter TPZ). The ionised shell of small planetary nebulae would (according to this model) result from the simple photoionisation of a high mass-loss rate stellar wind. IC 418 may be such a small nebula. The angular radius of the ionised nebula of IC 418 is $\theta_{ion}$ = 7.5" (Balick 1987) while the nebula's assumed distance is $D$ = 1 kpc, following Taylor & Pottasch (1987). The linear radius of the ionised nebula is therefore calculated to be $R_{ion} \approx 0.04\,\mathrm{pc}$, sufficiently small that the photoionised wind model may be accepted as valid for this nebula (GT). It has also been shown (TPZ) that the radio continuum spectrum of this nebula is well represented by the photoionised wind model. The similarity of the expansion velocities of the HI and ionised gas in this nebula also suggests that a photoionised wind model is appropriate (TGP). A simple photoionisation of a stellar wind is therefore used to model the growth of the nebula's ionised region to the present time. Although the model has also been run to simulate evolution to times well past the nebula's current age, this was done for reasons of academic curiosity only and it is acknowledged that a simple photoionised wind model is likely to be inadequate to model the evolution of the ionised nebula beyond a radius of $R_{ion} \approx 0.05\,\mathrm{pc}$.

The AGB wind is assumed to have a constant velocity and mass-loss rate for all times until the temperature of the star reaches $T_{AGB}$ = 6000 K, at which time the wind ends abruptly. The AGB wind consequently becomes detached from the stellar photosphere, creating an evacuated region between the stellar photosphere and the wind's inner edge. The wind therefore has an inverse-square density power law between its inner and outer edges, but has zero density outside these boundaries.

Let $R_{inn}$ and $R_{ion}$ respectively represent the inner and outer edges of the ionised AGB wind material. The inner edge of the HII region at time $t$ is given by

$$R_{inn}(t) = v_{exp}(t - t_{AGB}) + R_{AGB} \tag{1}$$

where $t_{AGB}$ is the time of the end of AGB mass loss, $R_{AGB}$ is the stellar radius at the end of AGB mass loss, and where $v_{exp}$ is the AGB wind's expansion velocity.

Assuming spherical symmetry, the total number of ions in the HII region is

$$N_{ion}(t) = \left(\frac{\dot{M}}{\mu_{ion} m_H v_{exp}}\right)\left(R_{ion}(t) - R_{inn}(t)\right) \tag{2}$$

where $\mu_{ion}$ is the mean molecular weight of the ionised gas, where $m_H$ is the weight of the hydrogen atom, and where $\dot{M}$ is the mass-loss rate.

Let $\dot{UV}(t)$ be the number of ionising photons produced by the star every second. Assume that there is no absorption between the star and $R_{inn}$ so that $\dot{UV}(t)$ also equals the number of photons impacting the inner edge of the wind material. Also assume that the nebula is ionisation bounded. Then $\dot{UV}(t)$ is also equal to the number of ionisation events per second:

$$\dot{UV}(t) = \dot{N}_{ion}(t) \tag{3}$$.

The increase in the mass of the ionised shell is determined by the excess of the ionisation rate over the recombination rate. The number of ions added to the shell per second equals the mass flux across the outer edge of ionised shell $R_{ion}(t)$. Let the number of new ions produced per second be equal to $\dot{N}_{ion(new)}(t)$. Then

$$\dot{N}_{ion(new)}(t) = \left(\frac{\dot{M}}{\mu_{neu} m_H v_{exp}}\right)\left(\frac{dR_{ion}}{dt} - v_{exp}\right) \tag{4}$$

for a mean molecular weight of the neutral gas of $\mu_{neu}$.

The ions are destroyed by the process of radiative recombination to the second or higher electronic energy level. The rate of recombinations per unit volume is given by the product of the number density of electrons $n_e$, the number density of ions $n_{ion}$, and $\alpha_2$, the reaction rate coefficient of radiative recombinations that result in a hydrogen atom in the second or higher energy level. The total rate of ion recombination is simply this product integrated over the volume of the ionised shell:

$$\dot{N}_{ion(old)}(t) = \int_{R_{inn}(t)}^{R_{ion}(t)} 4\pi\, n_{ion}(r)\, n_e(r)\, \alpha_2(r)\, r^2\, dr \qquad (5)$$

Let $n_{ion} = \gamma n_e$ and replace $n_e$ in equation 5 with the expression for $n_e$ suitable for a constant mass-loss rate model. Assume that the nebula is isothermal so that $\alpha_2 \neq \alpha_2(r)$. Also assume that the density of electronically excited hydrogen is negligible so that hydrogen ionisation by photons within the Balmer continuum or with lower energy can be neglected. Integration of equation 5 then yields

$$\dot{N}_{ion(old)}(t) = \frac{\alpha_2\, \gamma}{4\pi} \left( \frac{\dot{M}}{\mu_{ion}\, m_H\, v_{exp}} \right)^2 \left( \frac{1}{R_{inn}(t)} - \frac{1}{R_{ion}(t)} \right) \qquad (6)$$

Assuming a black-body radiation source for the star with radius $R_*(t)$ and temperature $T_*(t)$, the growth of the ionised region can be solved by combining equations 3, 4, and 6 to yield

$$\frac{dR_{ion}}{dt} = \begin{cases} v_{exp} \\ -\dfrac{\alpha_2\, \gamma}{4\pi} \left( \dfrac{\dot{M}\, \mu_{neu}}{m_H\, v_{exp}\, \mu_{ion}^{\,2}} \right) \left( \dfrac{1}{R_{inn}(t)} - \dfrac{1}{R_{ion}(t)} \right) \\ + 8\pi^2 \left( \dfrac{R_*(t)}{c} \right)^2 \left( \dfrac{\mu_{neu}\, m_H\, v_{exp}}{\dot{M}} \right) \displaystyle\int_{\nu_0}^{\infty} \dfrac{\nu^2}{e^{h\nu/kT_*(t)} - 1}\, d\nu \end{cases} \qquad (7)$$

Equation 9 is a first order differential equation of the form $\dfrac{dy}{dx} = f(x,y)$. It can be solved by numerical integration provided values of stellar radius $R_*(t)$ and stellar temperature $T_*(t)$ are available. The values of $R_*(t)$ and $T_*(t)$ are taken from Schönberner (1983) who determined theoretical $R_*(t)$ and $T_*(t)$ tables for planetary nebula nuclei (PNNi) with four

14masses: $M_* = 0.644\,M_\odot$, $M_* = 0.598\,M_\odot$, $M_* = 0.565\,M_\odot$, and $M_* = 0.546\,M_\odot$. The equations 1 and 7 can therefore be used to solve for $R_{inn}(t)$, and $R_{ion}(t)$, respectively.

The ionised gas expansion velocity of IC 418 is taken to be $v_{exp} = 17\,\text{km s}^{-1}$ from Weinberger (1989). As mentioned, the linear radius of the ionised region of IC 418 is assumed to be $R_{ion} \approx 0.04\,\text{pc}$. It is therefore required that the solution for the inner edge of the ionised region, given by equation 1, be $R_{inn} < 0.04\,\text{pc}$. Therefore, in order for $R_{inn}$ to be less than 0.04 pc, the time interval between the end of AGB mass loss (at a temperature of $T_{eff} = 6000\,\text{K}$) and the present time must be $\Delta t \leq 2100\,\text{yr}$, otherwise the expansion of the inner edge (which grows at a velocity of $v_{exp} = 17\,\text{km s}^{-1}$) would have taken it beyond the observed angular size of the *outer* edge. The time limit of $\Delta t \leq 2100\,\text{yr}$ places serious restraints on the evolutionary time scale of the central star, as it is required that the star must reach the observed temperature of IC 418's central star within this time period. The effective temperature of the central star of IC 418 is $T_{eff} = 33000\,\text{K}$ (Kaler 1976) and the only Schönberner evolutionary track that progresses from $T_{eff} = 6000\,\text{K}$ to $T_{eff} = 33000\,\text{K}$ in $\Delta t \leq 2100\,\text{yr}$ is the $M_* = 0.644\,M_\odot$ track. This track is consequently used in the model.

An original stellar mass must also be assumed for the star since this mass determines the total mass of material available for ionisation. An original zero-age-main-sequence (ZAMS) mass of $M_{*(ZAMS)} = 8\,M_\odot$ is chosen, based on the probable ZAMS mass of a planetary nebula with a $0.644\,M_\odot$ central star, resulting in a total mass of lost wind material of $M_{gas} = 7.356\,M_\odot$. The lack of an observed CO component in the neutral envelope of IC 418 and comparison with other nebulae (e.g. Likkel et al. 1992) leads us to believe that IC 418 is in fact not an exceptionally massive object, and so the assumed original stellar mass is likely to be an overestimate. The actual value of the initial mass is however unimportant because it is only the mass-loss rate and not the total mass that determines local gas densities. The total mass of the star is used only to calculate the mass of the reservoir of $H_2$ available for photodissociation. For this purpose, it is best to make sure that the reservoir of $H_2$ sufficiently large that there is no danger of creating an envelope with an outer edge smaller than its observed size for all of the mass-loss rates under consideration. The high ZAMS stellar mass assumed in the model accomplishes this, but does not otherwise affect the model.

A number of photoionised wind calculations (equations 1 and 9) have been performed with a variety of mass-loss rates in order to determine the best model mass-loss rate by comparison of the subsequent model value of $R_{ion}$ to the observed value of $R_{ion}$ = 0.04 pc at the time when the model star has reached the observed effective stellar temperature of $T_{eff}$ = 33000 K. The results are shown in figure 6, where it is seen that a mass-loss rate of $\dot{M}$ = 5×10$^{-5}$ $M_\odot$ yr$^{-1}$ results in the closest agreement with the observed value of $R_{ion}$ for the assumed stellar evolutionary track. This mass-loss rate is also consistent with that found by analysis of the radio continuum spectrum of the nebula by TPZ (after scaling to the assumed distance of 1 kpc). It is this mass-loss rate that was subsequently used in the modelling of the photodissociation region.

V. MODEL OF THE DISSOCIATION OF CIRCUMNEBULAR HYDROGEN MOLECULES

Computer modelling of $H_2$ photodissociation by stellar ultraviolet radiation has been conducted by Roger & Dewdney (1992; hereafter RD). The model is described in detail in RD, so only a cursory description of this model and a description of the modifications made to this model for the purposes of the present paper will be given here.

The original RD model traces the dissociation and subsequent ionisation of molecular gas surrounding a newly formed hot star. The main variables of the model are the effective temperature of the star and the density of the gas, the latter of which may be uniform or decline with distance according to an arbitrary power-law density gradient. Spherical symmetry is assumed. No dissociation of molecular hydrogen occurs by any source other than by the Lyman and Werner band photons produced by the central star. The inner HII region is assumed to be completely opaque to ionising radiation but completely transparent to Lyman-Werner photons. The population levels of the various Lyman and Werner vibrational states are determined solely by local UV radiation energy densities, thus collisional excitation and deexcitation are ignored. Gas-phase reactions that form $H_2$ are not considered because temperatures and densities are assumed to be far too low. Therefore atomic hydrogen accumulates in the model cloud because of an imbalance in the rate of molecular dissociation and the formation of molecules by catalyzed reactions on the surfaces of dust grains.

The energy density of stellar ultraviolet photons decreases with increasing distance from the star as required by the inverse square law, but also because of dust absorption and scattering, and because of the line optical depths of the Lyman and Werner bands. The extinction of ultraviolet radiation due to dust particles is calculated from the column density of gas by using an assumed ratio between these parameters. The subsequent (diluted and attenuated) UV energy density determines the populations of the various vibrationally and electronically excited states of the $H_2$.

The molecular gas is assumed to be heated by four mechanisms: the dissociation of $H_2$; the photoejection of electrons from dust grains; the ionisation of carbon atoms; and ionisation by cosmic rays. The gas is cooled by thermal excitation and subsequent re-radiation of energy by CI, CII, OI, FeII, and SII via line emission at infrared wavelengths as discussed by Dalgarno & McCray (1972). Quadrupole radiation produced by the $J = 2 \to 0$ $H_2$ rotational transition is also considered as a cooling process, as are various vibrational-rotational transitions considered by Hartquist, Oppenheimer, & Dalgarno (1980). Dipole rotational radiation by carbon monoxide transitions is not considered since IC 418 has not been observed as a CO radiation source (Huggins & Healy 1989) and it is consequently assumed that the CO is completely dissociated.

The model calculations are made as functions of both radius and time. The program calculates the density of atomic material outward from the outer edge of the Strömgren sphere inside a program loop, with each step of the loop used to determine the HI density at a slightly larger radius. The spacings of the radius points are variable; with points being more closely spaced at those radii where it is determined that the HI density changes most rapidly. The radius loop progresses outward until no significant HI density is found. The model is then progressed through a small time step. The HI density as a function of radius is then calculated again and the process is repeated.

To model the geometry and evolution of the neutral envelope of a planetary nebula, three major changes were made to the original RD model: the introduction of an evolving central star; the introduction of a detached and outwardly expanding shell of gas with a finite inner and outer radius; and the introduction of a HII region growing in a manner similar to an evolving planetary nebula, as determined by the photoionised wind model presented in the previous section.

Another required modification to the program involves the dust content of the gas. Circumstellar envelopes are dustier than the general interstellar medium, making it probable that a similar column density of gas in the planetary nebula model would produce much more extinction than in the original RD model, which was designed for young stars within unmodified interstellar material. In the case of IC 418, a HI column density of $\eta_{HI} = 2.9\times10^{23}\,\mathrm{m^{-2}}$ (TGP) corresponds to a local reddening of the ionised nebula of $E_{B-V} = 0.20$ and a visual extinction of $A_V = 0.62$ (Phillips, 1984). Assuming that all of the envelope is atomic, a one magnitude column density of $\eta_{mag} \approx 5\times10^{23}\,\mathrm{m^{-2}}$ would be derived. However, the assumption of a purely atomic envelope is not realistic, so a slightly higher one-magnitude column density of $\eta_{mag} = 1.0\times10^{24}\,\mathrm{m^{-2}}$ is therefore adopted.

The heavy element abundances of IC 418 are taken from Aller & Czyzak (1985) and are used to replace the helium, oxygen, carbon, and sulphur abundances assumed in the original RD model. No published determination of the iron and silicon abundance in IC 418 could be found, so solar abundances (de Jong, 1977) are assumed for these elements. Fortunately the model is quite insensitive to the adopted abundance of iron and silicon. A change of a factor of ten in these abundances creates only a few percent change in the radius of the HI shell. The adopted chemical abundances are listed in table 2.

The RD model assumes that the molecular gas possess a turbulent velocity of $\Delta v = 4\,\mathrm{km\,s^{-1}}$. This turbulence serves to broaden the Lyman and Werner absorption lines, effectively making the H$_2$ more susceptible to dissociation. However, a different assumption is made for the planetary nebula model. It is assumed that the stellar wind is non-turbulent (laminar), and that the Döppler width of the absorption lines is determined by the molecules' thermal velocity, given by the Maxwell-Boltzmann relation.

Models by Schönberner (1983) of the temperature and luminosity of planetary nebulae nuclei as a function of time are incorporated into the model. At each time step the ultraviolet energy density appropriate for the new stellar temperature and luminosity is calculated. A black-body energy distribution is assumed in the determination of the ultraviolet energy distribution.

The radius of the outer edge of the AGB wind is determined by the mass of material within the circumnebular envelope (that is, the amount of material lost by the AGB wind) which is determined by the difference in the assumed initial (main sequence) and final masses of the central star.

The model envelope is allowed to expand outward for a short period after the end of mass loss, during which time no dissociation is assumed to occur. This must be done to allow the gas density at the inner edge of the wind material to decrease to the point where the program can successfully run without any infinite values being generated by the calculations. Also, the gas density must be $n \lesssim 10^4\,\text{cm}^{-3}$ in order for the assumption that there is no collisional excitation to be valid. The required delay between the end of mass loss and the start of dissociation calculations is however fortunately quite brief. Dissociation calculations could successfully begin while the model central star was still very cool ($\lesssim 7000\,\text{K}$), before which time no significant dissociation or ionisation is expected to occur anyway. However, a potentially much more problematic assumption must also be made; that the gas is not clumpy, since significant inhomogeneities may lead to densities of $n > 10^4\,\text{cm}^{-3}$. The addition of clumps to the model is left as a non-trivial enhancement for the future.

The inner and outer edges of the HII region ($R_{inn}(t)$ and $R_{ion}(t)$, respectively) are determined at each point in time by the separate model calculations discussed in the previous section. The determined values of $R_{inn}(t)$ and $R_{ion}(t)$, as well as the current temperature and luminosity of the star as determined by Schönberner (1983), are then used in the dissociation model. The density of HI is then calculated outward up to the radius at which no appreciable HI is found, or at the outer edge of the neutral envelope, whichever comes first. Also calculated and stored is the radius at which one half the molecules are dissociated, which is defined to be the dissociation radius, $R_{HI}(t)$.

The model ionised nebula presented in the previous section has been incorporated into the $H_2$ dissociation code as a model HII region in order to determine if the observed mass of HI could plausibly have been formed by dissociation of molecular material by the ultraviolet radiation from the central star. The result is shown in figure 7.

The $H_2$ dissociation model predicts that by the time the star reaches an effective temperature of $T_{eff}$ = 33000 K, the mass of dissociated HI reaches $M_{HI} = 0.8 M_\odot$; much more than the mass of HI that the empirical model to the morphology of the envelope (above) would place in the inner HI shell alone, but similar to the mass of HI observed in the nebula as a whole (*i.e.* both the inner and outer HI shells together). The size of the model HI shell is $R_{HI} = 0.34$ pc; much larger than the empirical model's inner HI shell radius ($R_{HI} = 0.04$ pc) but significantly smaller than the observed value for the extent of the envelope of $R_{HI} \geq 0.4$ pc. The current photodissociation model therefore does not accurately reproduce the empirical model of the morphology of IC 418 except in its grossest feature of total HI mass.

It seems possible that a more sophisticated photodissociation model could bring the predicted and the observed photodissociation region into closer agreement. The addition of some interstellar dissociation into the model – which is expected at primarily larger radii – can easily increase the radius of the HI shell. It has already been shown (TGG) that the atomic gas at large radii could plausibly be the result of interstellar dissociation.

The much larger HI mass predicted by the predicted stellar photodissociation is less easily reconcilable with observations. However, the assumption of a black-body ultraviolet radiation source surely overestimates the flux of dissociating radiation. Since the rate of dissociation scales directly with the local density of UV radiation, the inclusion of a realistic stellar atmosphere model would reduce the accumulated HI mass. The substitution of the Kurucz (1979) stellar atmosphere models for the black-body distribution results in a small (10%) reduction in the accumulated HI.

The assumed mass-loss rate also has a profound effect on the accumulated mass of HI, with lower mass-loss rates resulting in lower masses of HI. This occurs because the reformation of molecules from constituent atoms is less efficient than the recombination of atoms from constituent electrons and protons, therefore the net ionisation rate is decreased more by increasing the gas densities than is the net dissociation rate. Higher mass-loss rates models therefore have more accumulated HI than low mass-loss rate models. It must therefore be considered possible that the assumed mass-loss rate is too high, even though the adopted mass-loss rate is consistent with that derived from radio continuum observations of TPZ.

Another model assumption that may result in an excessive amount of accumulated HI is the restricted number of cooling mechanisms included in the model. Specifically, cooling by dipole radiation from molecular rotational transitions are not included. Although no such radiation has yet been observed in IC 418, these processes are very efficient cooling mechanisms of warm gas. If the gas temperature were lowered by the inclusion of these processes, the Döppler widths of the $H_2$ absorption lines would be correspondingly reduced. Test runs with different Döppler line widths indicate that the accumulated HI mass scales approximately to the square root of the gas temperature.

The model of the underlying HII region must also be called into question, as it is a very simple photoionised wind model. However, the model HII region used has actually very little affect on the accumulation of HI in the model since the model HII region only describes the outer edge of the ionised shell (and hence the inner edge of the HI shell) as a function of time. The adopted photoionised wind model is in close agreement in this respect to various other models, including interacting stellar winds models, within the current astronomical literature. For example, in a model planetary nebula by Kahn & Breitschwerdt (1990) an ionised radius of $R_{ion}$ = 0.05 pc is obtained at a time $t$ = 2000 yr after the photospheric temperature of the central star reaches $T$ = 5600 K, quite similar to the evolutionary time scale of the model presented here. However the Kahn & Breitschwerdt models also predict that significant acceleration of the ionised gas would have also occurred during this time, resulting in an ionised shell that expands $\approx$ 10 km s$^{-1}$ faster than the surrounding envelope. This is not in keeping with the observed kinematics of IC 418 or the lack of shock-excited H$_2$ emission observable in IC 418 (Webster et al. 1988), making the photoionised wind model our preferred model despite the two models' similar predictions concerning ionised radius.

The model photodissociation region seems to be far more similar to the atomic envelope observed in the nebula BD+30° 3639 than that of IC 418. The atomic envelope of BD+30° 3639 has a similar mass of $M_{HI}$ = 0.27 $M_\odot$ albeit a smaller radius of $R_{HI} \leq$ 0.16 pc (Likkel et al. 1992). The stellar photodissociation model therefore seems adequate for the atomic envelopes of some planetary nebulae, but not IC 418. It is not known why the HI cloud of IC 418 would be unusual in this regard, but differences in stellar mass and chemical composition are likely to be significant.

VI. CONCLUSION

There seems to be no difficulty in creating a significant mass of circumnebular HI surrounding a planetary nebula by molecular photodissociation by radiation from the central star. Therefore there is no apparent need to postulate additional sources of accumulated HI, such as an atomic wind in late or post-AGB evolution or by molecular dissociation from the passage of a shock. The observed geometry of the HI observed in IC 418 is however not accurately reproduced by the current model, although the model appears to be consistent with what is known about the much more poorly studied nebula BD+30° 3639.

## VII. ACKNOWLEDGMENTS

This research is supported by a grant from the Natural Sciences and Engineering Research Council of Canada. This research has made use of the SIMBAD database, operated by CDS, Strasbourg, France.

TABLE 1

Parameters of Best Empirical Models of the Atomic Envelope of IC 418 [1]

| | Inner HI Shell | Outer HI Shell |
|---|---|---|
| Inner Radius | 7.5" ± 1" | 25" ± 3" |
| Outer Radius | 9" ± 1" | ≥ 90" |
| HI Mass | 0.00 $M_\odot$ to 0.03 $M_\odot$ | 0.2 $M_\odot$ to 0.8 $M_\odot$ |
| HI Temperature | 10 K to 300 K | 100 K to 600 K |
| Mass-Loss Rate (as atoms): | $5\times10^{-6} M_\odot$ yr$^{-1}$ to $2\times10^{-5} M_\odot$ yr$^{-1}$ | |
| Expansion Velocity: | 17 km s$^{-1}$ ± 1 km s$^{-1}$ | |
| Gas Density Power Law | $n \propto r^{-\alpha}$ where $\alpha \gtrsim 2$ | |

[1] For an assumed distance of 1 kpc.

TABLE 2

Model Elemental Abundances

| Element | Abundance | Element | Abundance |
|---|---|---|---|
| Helium | $\frac{n_{He}}{n_H} = 9.3 \times 10^{-2}$ | Sulphur | $\frac{n_S}{n_H} = 4.2 \times 10^{-6}$ |
| Oxygen | $\frac{n_O}{n_H} = 4.36 \times 10^{-4}$ | Silicon | $\frac{n_{Si}}{n_H} = 3.5 \times 10^{-5}$ |
| Carbon | $\frac{n_C}{n_H} = 6.2 \times 10^{-4}$ | Iron | $\frac{n_{Fe}}{n_H} = 2.5 \times 10^{-5}$ |

FIGURE CAPTIONS

FIGURE 1. Map of circumnebular atomic hydrogen emission from VLA D configuration data. The size of the cross indicates the synthesised beam size, and the cross position indicates the center of the ionised nebula.

FIGURE 2. $\lambda = 21\,\mathrm{cm}$ spectrum of IC 418 from VLA D configuration data.

FIGURE 3. $\lambda = 21\,\mathrm{cm}$ spectral line flux within concentric annuli of 10" width as a function of radius.

FIGURE 4. A $z = 0$ cross section of a spherically-symmetric model of an ionisation-bounded planetary nebula (not necessarily to scale) showing the central star (black), the central evacuated hollow (white), the dense ionised region (dark grey), the molecular region (medium grey), and the atomic regions (light grey). Directions follow the usual Cartesian convention, with positive $x$ to the right and positive $y$ upwards. The $z$ axis (not shown) is positive out of the page.

FIGURE 5. Best fit empirical models using four model morphologies. The models are: a constant gas density ($n \propto r^0$) and constant expansion velocity model shown by a solid line; a constant mass-loss rate and a constant expansion velocity ($n \propto r^{-2}$) model shown by a dashed line; a constant mass-loss rate, radially increasing expansion velocity model ($n \propto r^{-3}$) shown by a dotted line; and a increasing mass-loss rate, constant expansion velocity model ($n \propto r^{-3}$) shown by a dash-dotted line. The best simultaneous fit to both the radial flux profile (5a) and the spectrum (5b) comes from the $n \propto r^{-2}$ model incorporating a constant expansion velocity and a constant mass-loss rate.

FIGURE 6. The inner and outer radii of the ionised region of model planetary nebulae ($R_{inn}$ and $R_{ion}$, respectively) as a function of time since the end of AGB mass loss, assumed to occur when $T_{eff} = 6000\,\mathrm{K}$. The numbered curves are the calculated values of $R_{ion}(t)$ for nebulae with the labelled mass-loss rates in units of $10^{-5}\,M_\odot\,\mathrm{yr}^{-1}$. The horizontal dashed line is the radius above which the photoionised wind model is believed to become unreliable

($R_{ion}$ = 0.05 pc). Model radii above this value are not believed to necessarily reflect those of real planetary nebulae, but are included for the sake of completeness. The vertical dashed line is that time at which the effective temperature of the central star is $T_{eff}$ = 33000 K, the observed temperature of the central star of IC 418. The best fit to the observed value of $R_{ion}$ = 0.04 pc for IC 418 is the curve for $\dot{M} = 5 \times 10^{-5}\ M_\odot\ \text{yr}^{-1}$.

FIGURE 7. Model radii (top) and masses (bottom) of the molecular, inner atomic, and ionised shell as a function of time since the end of AGB mass loss, assumed to occur when $T_{eff}$ = 6000 K. The model uses an assumed mass-loss rate of $\dot{M} = 5 \times 10^{-5}\ M_\odot\ \text{yr}^{-1}$ as determined by the best-fit model to the observed size of the ionised nebula (figure 6). The horizontal dashed line in the upper plot is the radius above which the photoionised wind model is believed to be invalid ($R_{ion}$ = 0.05 pc). The vertical dashed line is that time at which the effective temperature of the central star is $T_{eff}$ = 33000 K, the observed temperature of the central star of IC 418.

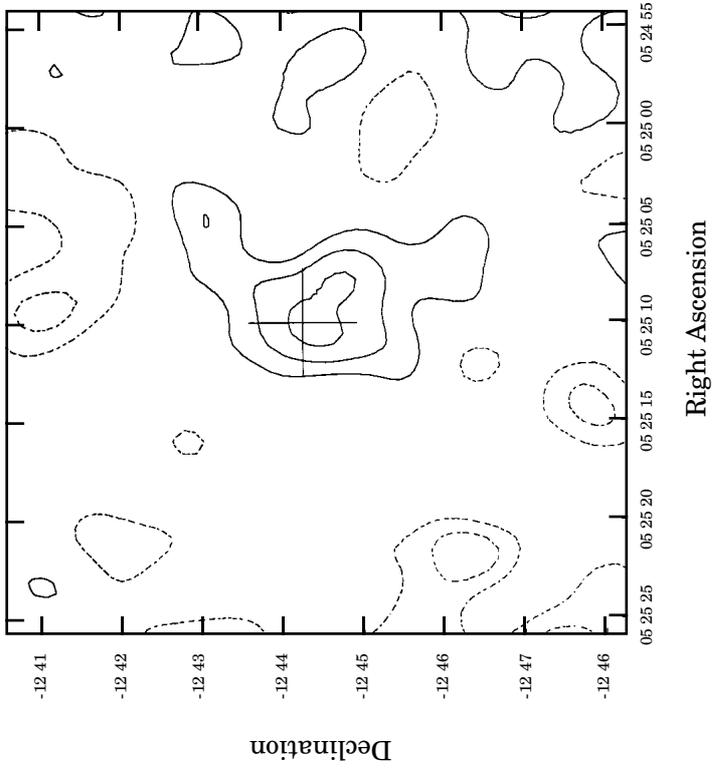

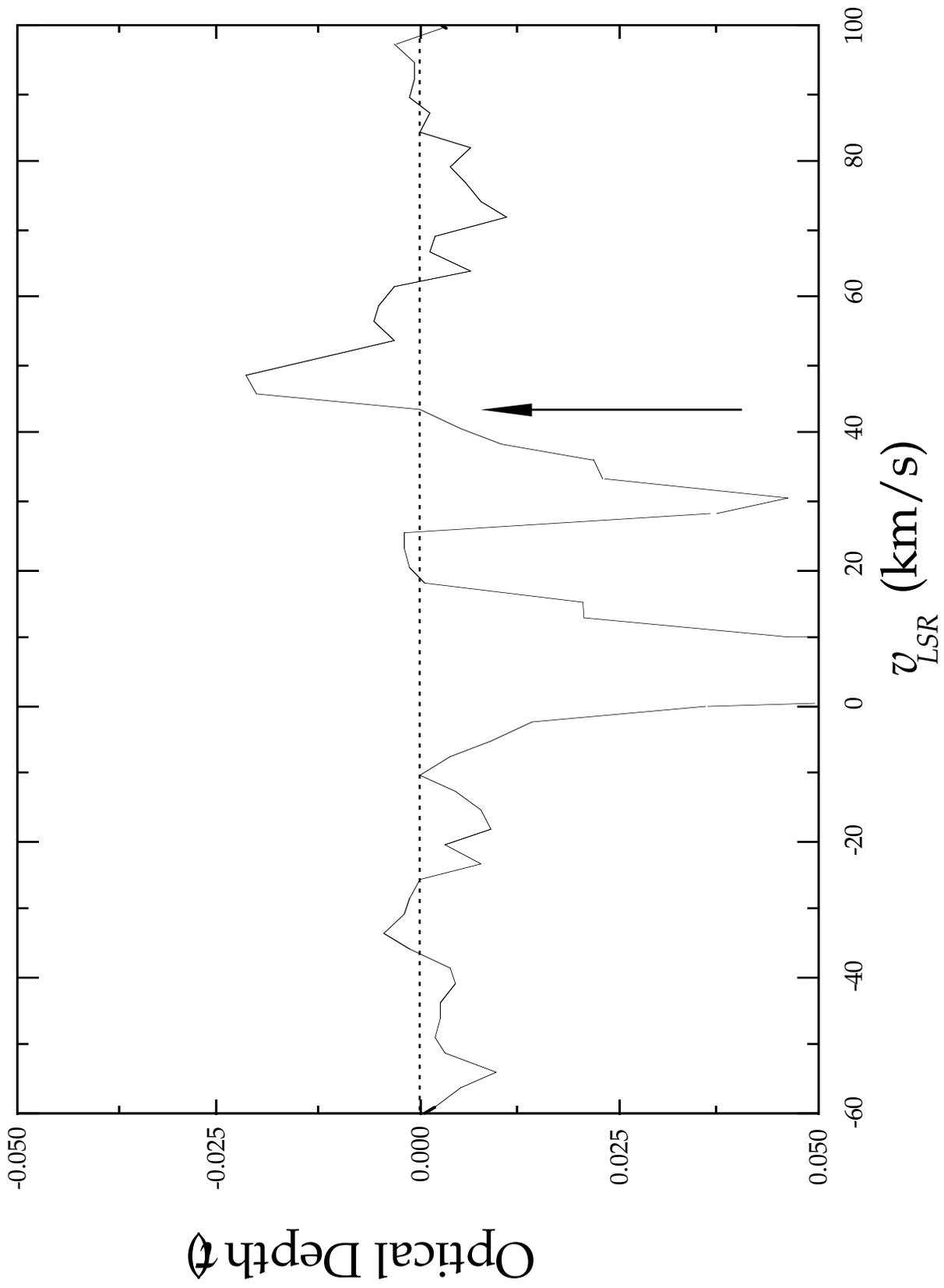

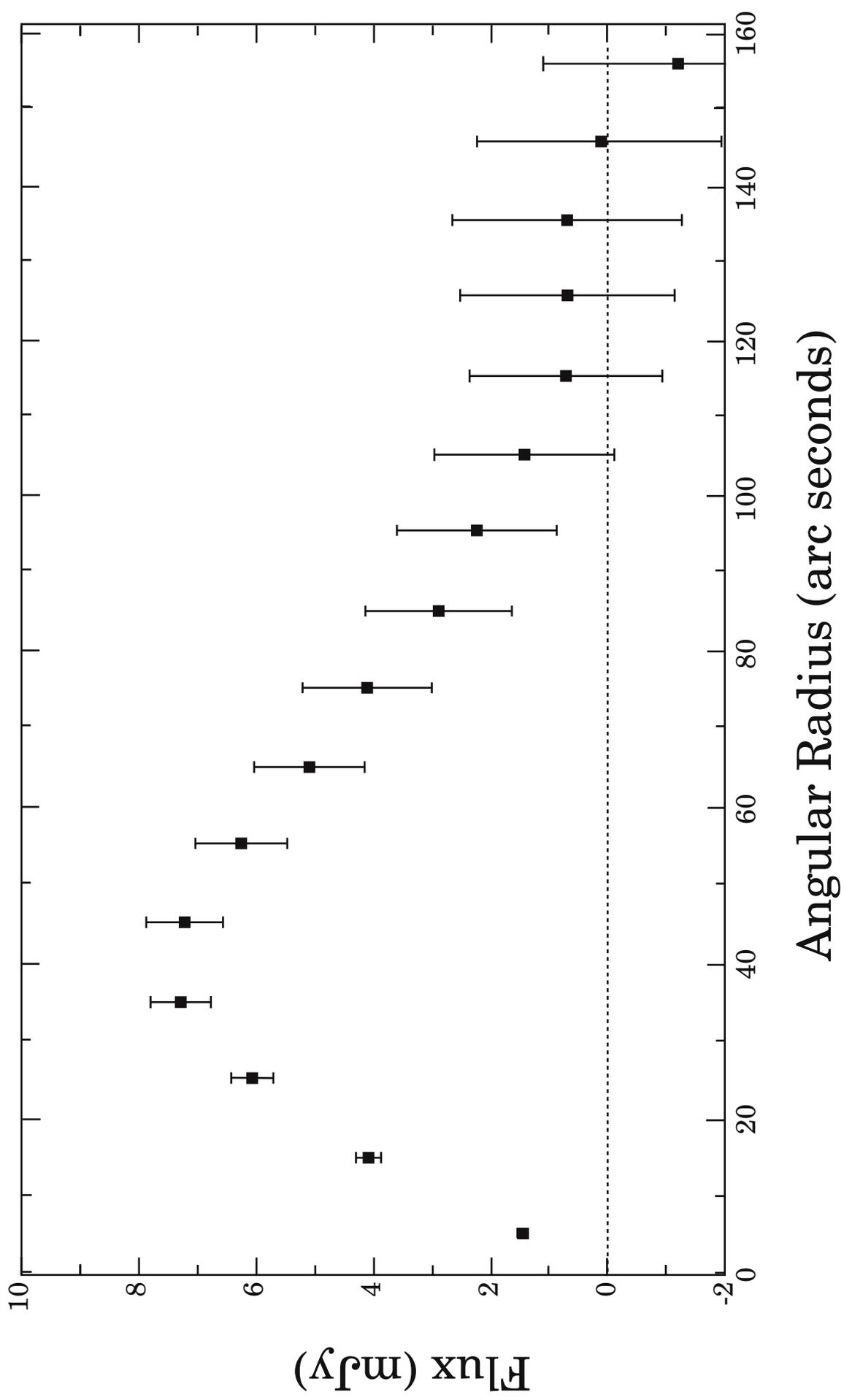

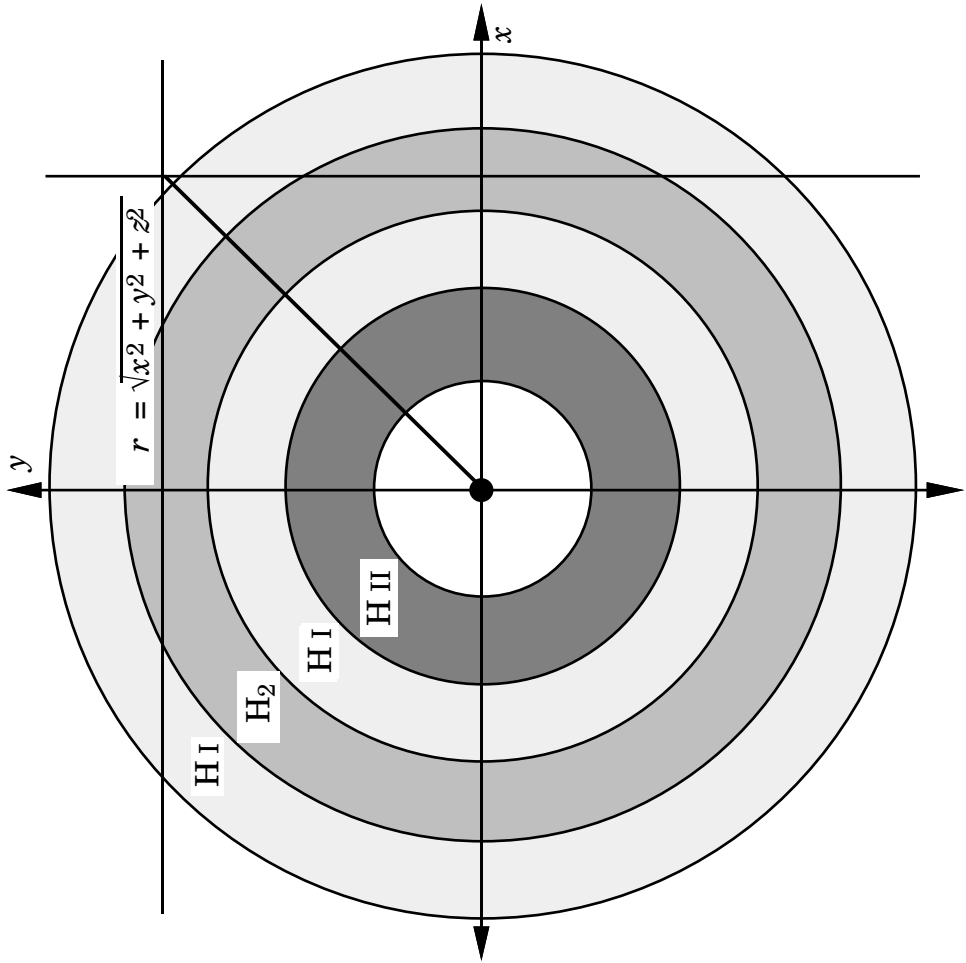

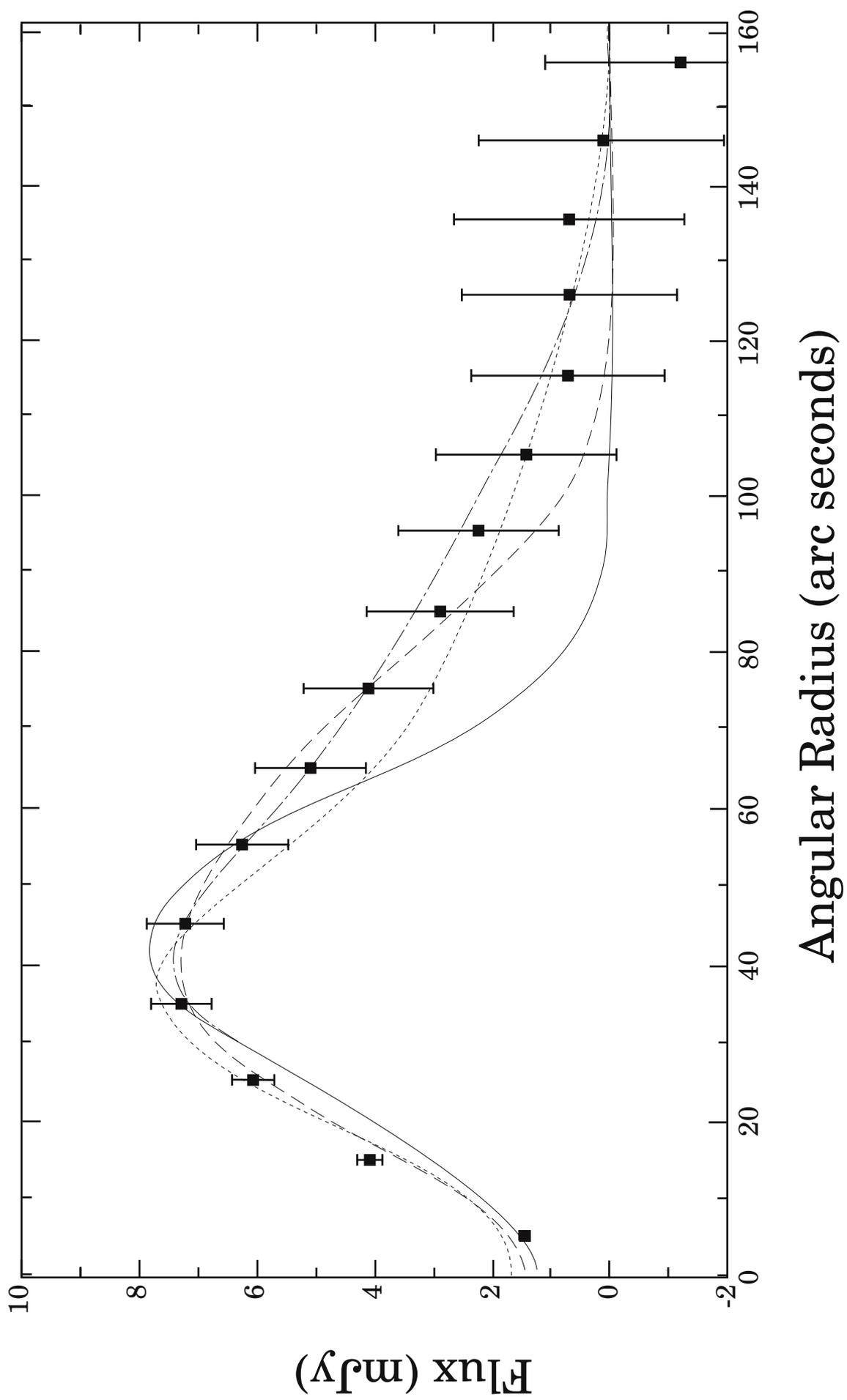

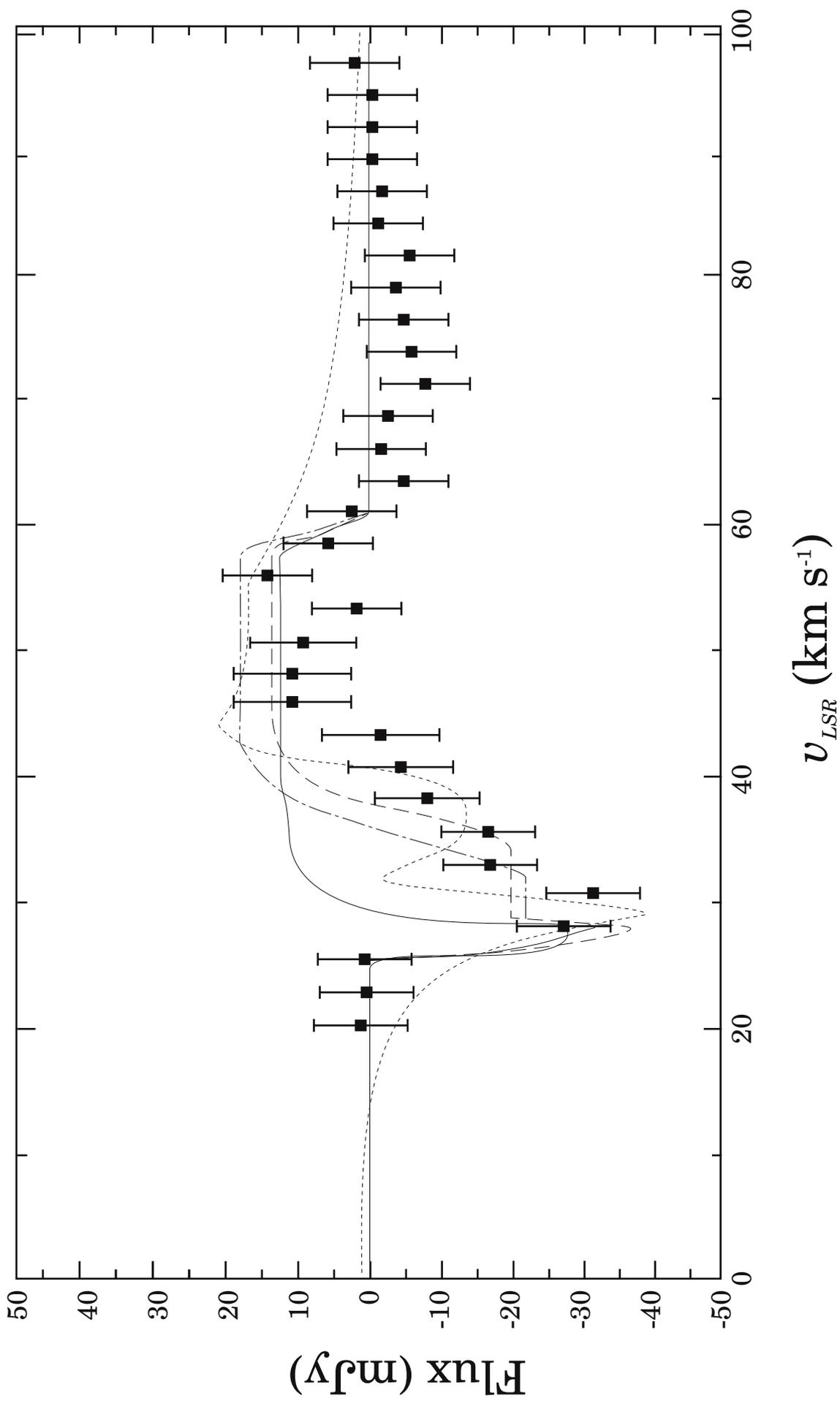

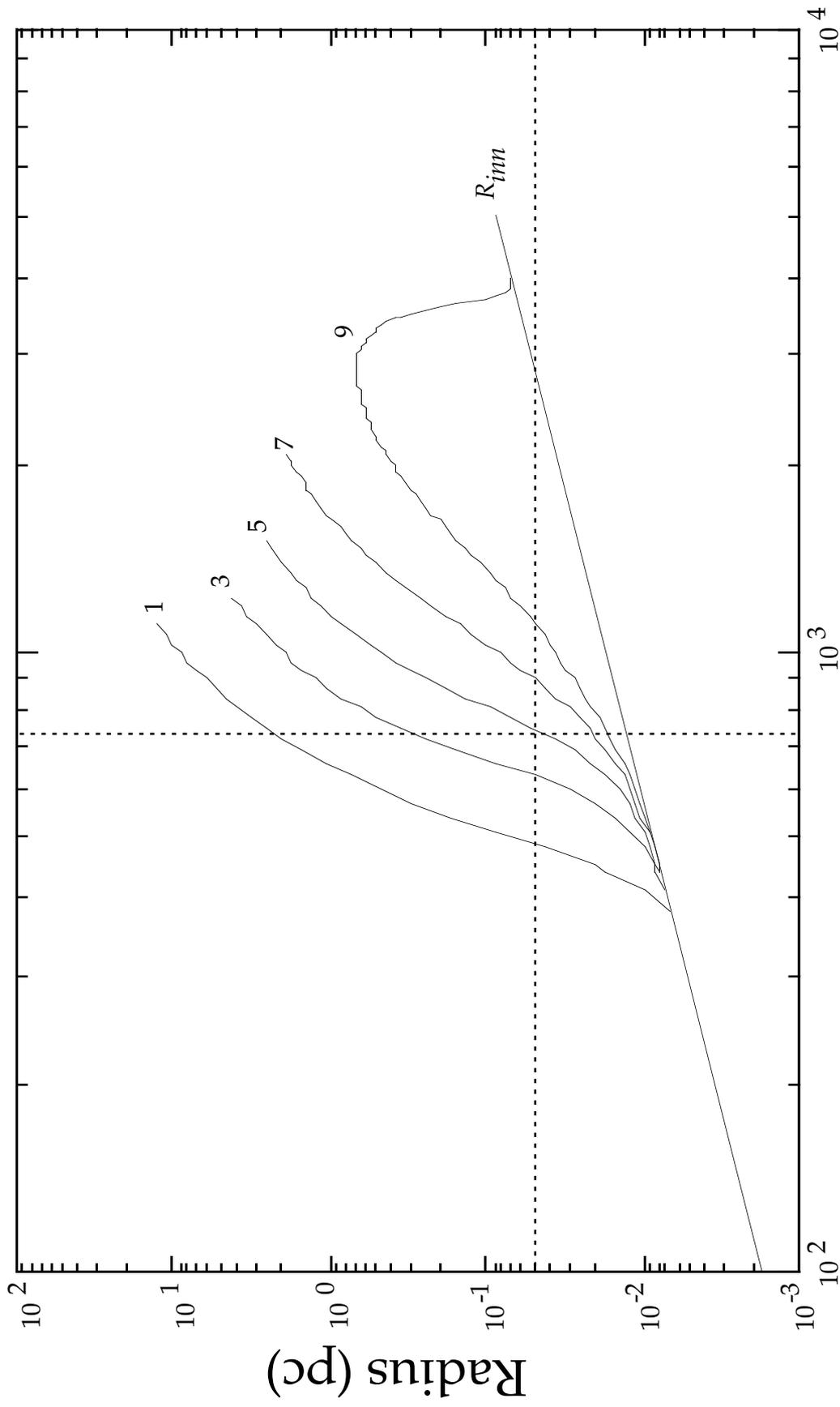

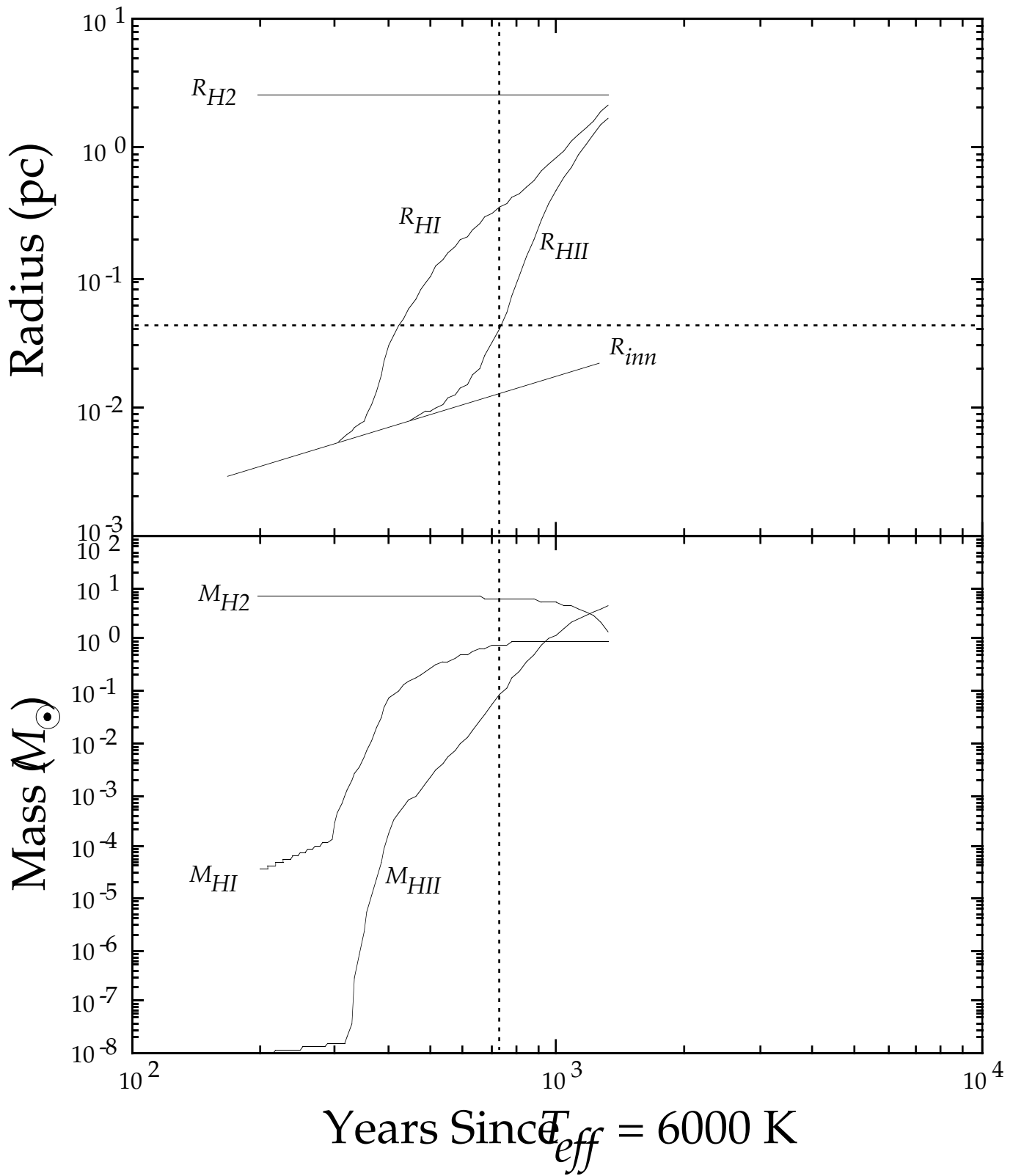